\documentclass[10pt, conference, letterpaper]{IEEEtran}
\IEEEoverridecommandlockouts
\usepackage[T1]{fontenc}
\usepackage[utf8]{inputenc}
\usepackage{textcomp}
\usepackage{microtype}
\usepackage{dblfloatfix}
\usepackage{amsmath,amssymb}
\usepackage{algorithmic}
\usepackage{graphicx}
\usepackage[caption=false,font=footnotesize,subrefformat=parens,labelformat=parens]{subfig}
\usepackage{xcolor}
\usepackage{verbatim}
\usepackage{pifont}
\usepackage{authblk}
\usepackage{siunitx}
\usepackage{xspace}
\usepackage{enumitem}

\newcommand{\simpletitle}[1]{\noindent\textbf{#1}.\xspace}

\newcommand{\toolname}{waveSLAM\xspace}

\usepackage[authormarkup=superscript,deletedmarkup=sout,addedmarkup=em]{changes}
\usepackage{soul}
\soulregister\cite7
\soulregister\ref7
\soulregister\pageref7

\colorlet{soulcyan}{cyan!30}
\colorlet{soulgreen}{green!30}

\definechangesauthor[name={Claudio}, color=green!75!black]{CF}

\usepackage{xurl}
\DeclareMathOperator*{\argmax}{arg\,max}

\usepackage{glossaries}
\newacronym{aoa}{AoA}{Angle Of Arrival}
\newacronym{ap}{AP}{Access Point}
\newacronym{cots}{COTS}{Commercial off-the-shelf}
\newacronym{ftm}{FTM}{Fine Timing Measurement}
\newacronym{tof}{ToF}{Time of Flight}
\newacronym{ura}{URA}{Uniform Rectangular Array}
\newacronym{ros1}{ROS-1}{Robot Operating System version 1}
\newacronym{mmwave}{mmWave}{millimeter-Wave}
\newacronym{mimo}{MIMO}{Multiple-In Multiple-Out}
\newacronym{isac}{ISAC}{Integrated Sensing and Communications}
\newacronym{jcas}{JCAS}{Joint Communications and Sensing}
\newacronym{slam}{SLAM}{Simultaneous Localization and Mapping}
\newacronym{sta}{STA}{Station}
\newacronym{csi}{CSI}{Channel State Information}
\newacronym{wlan}{WLANs}{Wireless Local Area Networks}
\newacronym{waveslam}{waveSLAM}{mmWave-based Simultaneous Localization and Mapping}

\title{waveSLAM: Empowering Accurate Indoor Mapping Using Off-the-Shelf Millimeter-wave Self-sensing}
\begin{document}
\author[1]{Pablo Picazo \thanks{This work has been partially funded by the European Union’s Horizon Europe research and innovation program  under grant agreement No 101095759 (Hexa-X-II) and the Spanish Ministry of Economic Affairs and Digital Transformation and the European Union-Next Generation EU through the UNICO 5G I+D 6G-EDGEDT.}
\thanks{©2023 IEEE. Personal use of this material is permitted. Permission
from IEEE must be obtained for all other uses, in any current or future
media, including reprinting/republishing this material for advertising
or promotional purposes, creating new or redistribution to servers or lists, or reuse of any copyrighted
component of this work in other work}
\thanks{Published at IEEE VTC FALL 2023 (Hong Kong)}
}

\author[1]{Milan Groshev}
\author[2]{Alejandro Blanco}
\author[3]{Claudio Fiandrino}
\author[1]{Antonio de la Oliva}
\author[3]{Joerg Widmer}
\affil[1]{Universidad Carlos III de Madrid. Madrid, Spain, papicazo@pa.uc3m.es}
\affil[2]{The University of Edinburgh. Edinburgh, Scotland }
\affil[3]{IMDEA Networks Research Institute. Madrid, Spain}




\maketitle

\begin{abstract}
This paper presents the design, implementation and evaluation of \toolname, a low-cost mobile robot system that uses the millimetre wave (mmWave) communication devices to enhance the indoor mapping process targeting environments with reduced visibility or glass/mirror walls. A unique feature of \toolname is that it only leverages existing Commercial-Off-The-Shelf (COTS) hardware (Lidar and mmWave radios) that are mounted on mobile robots to improve the accurate indoor mapping achieved with optical sensors. The key intuition behind the \toolname design is that while the mobile robots moves freely, the mmWave radios can periodically exchange angle and distance estimates between themselves (self-sensing) by bouncing the signal from the environment, thus enabling accurate estimates of the target object/material surface. Our experiments verify that \toolname can archive cm-level accuracy with errors below $22$~cm and $20^{\circ}$ in angle orientation which is compatible with Lidar when building indoor maps.

\end{abstract}

\begin{IEEEkeywords}
Indoor mapping, mmWave sensing,60 GHz, self-sensing, beamsweeping, FTM, CSI.
\end{IEEEkeywords}

\section{Introduction}
\label{sec:intro}

\gls{slam}~\cite{slam-pro} is the process that provides a real-time estimation of a mobile robot's location and the construction of a map of the environment where the robot is moving. According to a report by Straits Research, the global \gls{slam} technology market is expected to grow from USD 226.7 million in 2021 to USD 9425.7 million by 2030, at a compound annual growth rate of 49.41\%~\cite{m4}. Such growth is attributed to the increasingly growing demand by the industrial sector for autonomous mobile robots to improve productivity, and logistics, and reduce the production costs~\cite{m3}. Recent proprietary approaches to create indoor maps (e.g., Google Maps Indoor, HERE Indoor maps, and Apple indoor maps) are human-driven, complex, and not suitable for industrial scenarios. A promising alternative is using autonomous robots to create indoor maps.

State-of-the-art mobile robot solutions use optical sensors such as LiDARs~\cite{lidar}, RGB cameras~\cite{r1},~\cite{r2} or stereo cameras~\cite{camera-2} and produce very accurate indoor maps~\cite{r3}. However, these optical sensors experience significant performance degradation in the presence of dust, fog, or smoke and their use is reduced in indoor environments with glass/mirror walls or insufficient light. While airborne obscurants like dust or insufficient light can occur in manufacturing situations, sight glass is used commonly in panels, lenses, and covers for manufacturing equipment. 

\begin{figure}
\centering%
\includegraphics[width=0.375\textwidth]{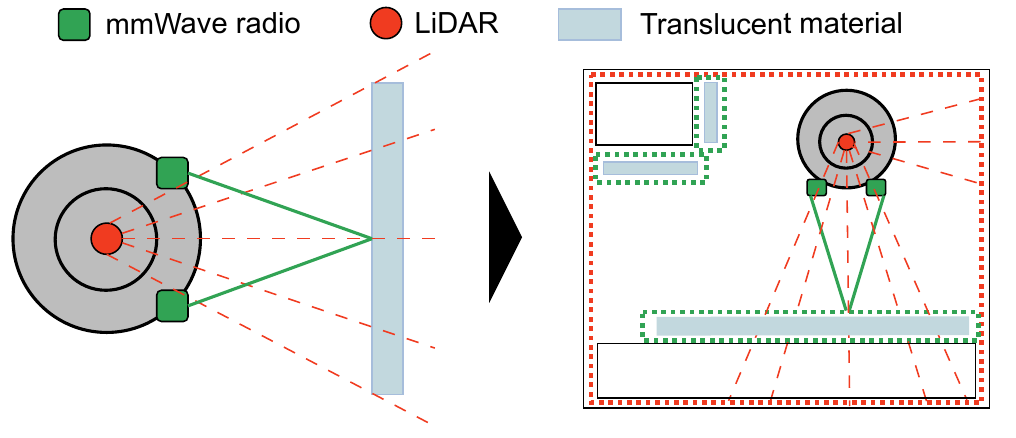}%
\vspace*{-2ex}%
\caption{Key intuition behind WaveSLAM}%
\label{fig:waveSLAM}%
\vspace*{-3ex}%
\end{figure}

The \gls{mmwave}-based sensing was introduced as a promising alternative solution for indoor map creation seen through the \gls{mmwave} radios~\cite{mwave-1,mwave-2,mwave-3,mwave-4,r10}. However, to enable practical, efficient, and accurate environment sensing, the existing solutions fall short in one of three major aspects: \emph{(i)} \textit{Access Points dependence} Due to the need for directional connectivity and the inherent spatial channel sparsity of 60 GHz devices, a single access point (AP) can only sense a limited part of the indoor environment~\cite{mwave-1,mwave-2}. As a consequence, existing solutions require multiple collaborating~\cite{mwave-4} or human interventions when sensing the indoor environment~\cite{mwave-1,mwave-2}. \emph{(ii)} \textit{Software-Defined Radios (SDR) dependence} Previous \gls{mmwave} sensing solutions~\cite{mwave-1,mwave-4} are prototyped using SDRs that are big and expensive and require from extra mixers to operate at mmWave\footnote{Available online (accessed 11/07/2023): \url{ https://www.pasternack.com/waveguide-mixer-down-converter-wr-12-60-90-ghz-if-18-ghz-pe12d1001-p.aspx}} which limits their practical applicability. \emph{(iii)} \textit{mmWave as a Sensor} Some recent efforts~\cite{r10} rely on radar-based solutions operating at \gls{mmwave} frequencies to overcome the shortcomings of LiDARs. However, such solutions are complex to integrate and incur high energy expenditure.

In this paper, we propose \toolname, an indoor mapping solution that exploits the sensing information available from 60 GHz mmWave \gls{cots} devices in mobile robots, without any external APs or network devices support. The main architectural approach we explore in this paper is the re-use of sensing information from the \gls{mmwave} devices already integrated into the mobile robot primarily for communication purposes to enhance the indoor map creation process when the optical sensors experience performance degradations (i.e. in case of LiDARs with translucent material). 

\toolname mounts LiDAR and two \gls{mmwave} radios on custom-grade commodity robots, which are used as environment sensors. The robot moves and scans freely the indoor environment using the LiDAR, while the \gls{mmwave} radios periodically exchange angle and distance estimates between themself (self-sensing) by bouncing the signal in the environment, thus enabling accurate estimates of the target object/material surface (see Fig.~\ref{fig:waveSLAM}). In Sec.~\ref{sec:experimental}, we show that \toolname can achieve cm-level accuracy which is compatible with LiDAR when building indoor maps. \textit{To the best of our knowledge, no prior work has proposed the use of \gls{cots} \gls{mmwave} self-sensing capability for improving the performance of existing optical sensor-based SLAM solutions.}

\simpletitle{Contributions} Indoor mapping and localization is a widely challenged problem still today. Unlike all the previous efforts, in this paper, we take a first step toward answering the question: \textit{How can we use the sensing information from \gls{cots} \gls{mmwave} radios available in existing mobile robots to improve exiting SLAM solutions?} We answer this question by proposing \toolname and we make the following main contributions:
\begin{itemize}
  \item We design \toolname, the first mmWave-based system that uses radio sensing and optical sensors to help create accurate indoor maps. To do so, \toolname leverages the \gls{ftm} procedure, to get distance and \gls{csi} data to derive \gls{aoa} estimates~(Sec.~\ref{sec:fundamentals}). These accurate estimates obtained from the mobile robot communication radios are combined with the mobile robot odometry and with LiDAR to produce a point cloud estimation (Sec.~\ref{sec:solution}).  
  \item We implement a real prototype of \toolname (Sec.~\ref{sec:prototype}) using \gls{cots} mmWave devices to extract \gls{csi} information for \gls{aoa} and \gls{ftm} ranging.  
  \item We evaluate \toolname through extensive experimental analysis (Sec.\ref{sec:experimental}) where we deployed our prototype in a real-world indoor environment. We compare \toolname to the traditional LiDAR approach in different scenarios and against different types of wall materials. Our results show that \toolname can achieve cm-level precision with errors below $22$~cm and $20^{\circ}$ in angle orientation across all configurations and settings.
\end{itemize}

\section{Principles of mmWave Localization}
\label{sec:fundamentals}
In this section, we explain the details of the \gls{mmwave} sensing techniques used in this work. We first describe the calculus of the \gls{tof} obtained via the \gls{ftm} procedure, we then explain the \gls{aoa} estimation with the \gls{ura}. 

\begin{figure}
\centering%
\subfloat[ToF extracted from FTM~\label{fig:ftm}]{%
    \includegraphics[width=.475\columnwidth,keepaspectratio]{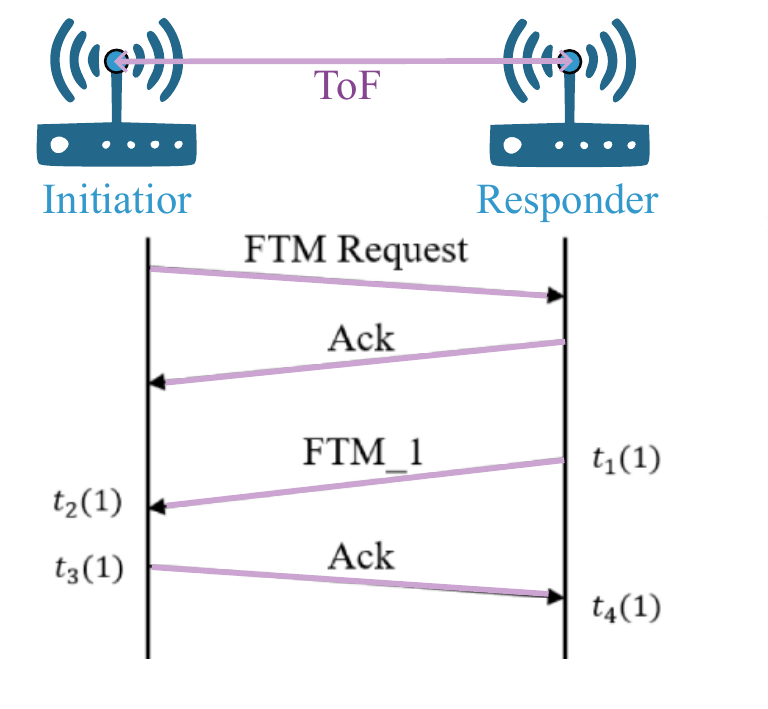}}%
~%
\subfloat[AoA estimations with URA~\label{fig:aoa}]{%
    \includegraphics[width=.475\columnwidth,keepaspectratio]{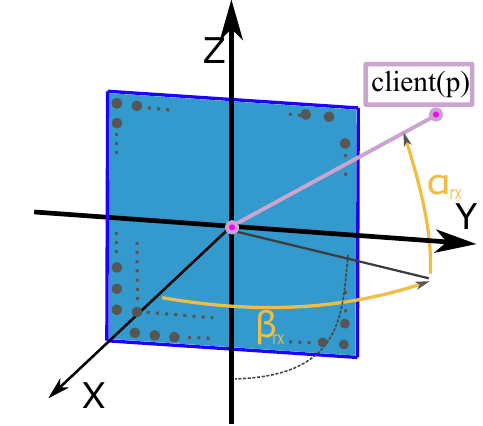}}%
\vspace*{-1ex}%
\caption{Principles of \gls{mmwave} localization}%
\label{fig:mmwave-principles}
\vspace*{-3ex}%
\end{figure}

\subsection{\gls{tof} over \gls{ftm} Procedure}
\label{sec:tof}
The IEEE P802.11-2016 describes the \gls{ftm} procedure that can be used by two \gls{ftm}-capable devices to estimate their distance without clock synchronization. The device that starts the \gls{ftm} session is the Initiator (AP), and the other is the Responder (STA). The Initiator starts the \gls{ftm} session by sending an \gls{ftm} request frame. The Responder then has $10$~ms to acknowledge the request and confirm that the \gls{ftm} session has started. An \gls{ftm} session contains one or more bursts that in turn are composed of one or more measurements. A burst is a specific period during which the channel is reserved for \gls{ftm} measurements so that all the measurements performed in the same burst are consecutive. Fig.~\ref{fig:ftm} shows an example with one burst consisting of 1 measurement. After the \gls{ftm} request has been acknowledged (Ack), the Responder sends the first \gls{ftm} packet with $t_1$ and $t_4$ set to 0. Then, for the rest of the measurements, the Responder sends an \gls{ftm} frame with the local clock timestamp of when it is sent ($t_1$), as well as a local clock timestamp of when the last Ack message was received ($t_4$). Thus, in \gls{ftm} frame $N$ we obtain information about the measurement $N - 1$ for $N > 1$. This is repeated until all measurements of the burst (max $N=32$) are done. 
We can estimate the \gls{tof} as follows:

 \begin{equation}
    \textrm{ToF} = \frac{1}{2n}\sum_{x=1}^{n} ((t_4(x) - t_1(x)) - (t_3(x) - t_2(x))),
    \label{eq:ftm}
\end{equation}
where $n$ is the number of measurements.

\subsection{\gls{aoa} Estimation with a \gls{ura}}
\label{sec:aoa}
Consider a wireless system where the transmitter and receiver have a single RF-chain connected to a \gls{ura} with $K\times J$ antenna elements. The signal sent by the transmitter propagates through the multipath channel along $P$ different paths and arrives at the receiver. Fig.~\ref{fig:aoa} exemplifies transmitter located at $(X_s,Y_s,Z_s)$ and the \gls{ura} lies in the $X = 0$ plane. We characterize the paths from the source as they appear at the center of the \gls{ura} using the following path parameters:

\noindent $\bullet$ \textbf{Complex attenuation} $\gamma$ of the signal over the path.\par
\noindent $\bullet$ \textbf{Elevation angle} $\alpha$ between the path and the plane $Z = 0$.\par
\noindent $\bullet$ \textbf{Azimuth angle} $\beta$ between the path and the plane $Y = 0$.

These path parameters are common for all the paths-see~\cite{multiloc}. We represent the multipath wireless channel (\gls{csi}) in terms of the path parameters as follows:
\begin{equation}
    H = \sum_{p = 1}^{P} H_p(\gamma_p, \alpha_p, \beta_p) + W .
\end{equation}
$W$ is white Gaussian noise. 

To separate the paths and extract the parameters, we follow the approach of mD-track~\cite{md-track}. This algorithm estimates the path parameters iteratively by reconstructing the strongest path and then subtracting it from the received channel so that successively weaker paths can be estimated. Note that mD-track was designed to work with a uniform linear array. The algorithm was later extended to adapt it to the \gls{mmwave} \gls{ura}~\cite{multiloc}.

After completing the mD-track algorithm, there are $P$ estimates of the path parameters and each one is associated with a path $p$. For the extracted $p$-th path, we denote $\hat{\gamma}_p$, $\hat{\alpha}_p$ and $\hat{\beta}$ as the estimated complex attenuation, the estimated elevation angle, and the estimated azimuth angle, respectively. In practice, the channel between Initiator and Responder is estimated by beam training using the beam-sweeping procedure. This consists of testing different beam patterns between the two \gls{sta}s.


\section{\toolname}
\label{sec:solution}

\begin{figure*}
\centering%
\includegraphics[width=0.95\textwidth]{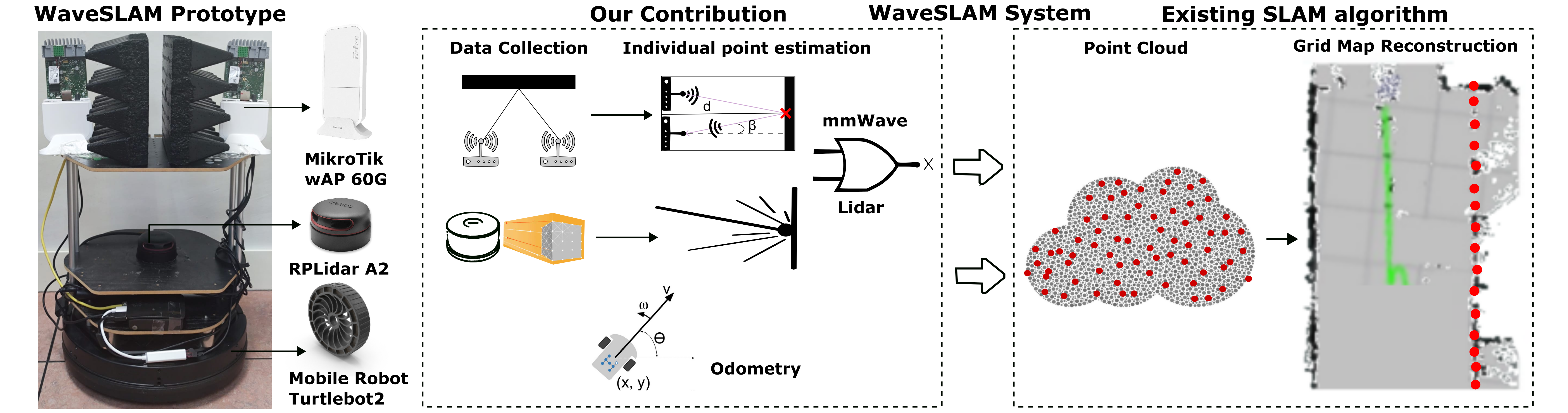}%
\vspace*{-2ex}%
\caption{WaveSLAM prototype--(left) and system building blocks--(right)}%
\label{fig:blocksys}%
\vspace*{-3ex}%
\end{figure*}

In industrial scenarios, robots are equipped with multiple sensors, i.e., visual cameras, LiDARs, and radio-frequency interfaces to communicate with their environment. A new generation of mobile robots may be equipped with multiple \gls{mmwave} antennas that can provide redundancy and increase reliability, resulting in higher communication rates and being able to send data even if one of the antennas is obstructed or fails. This section introduces \toolname, a solution that leverages \gls{mmwave} sensing to complement existing LiDAR-based SLAM solutions. Fig.~\ref{fig:blocksys}--(right) shows the basic building blocks of \toolname:


\subsection{Data Collection} This building block serves as a data collection frontend, by which \toolname senses the environment using the \gls{mmwave} devices and a LiDAR. \gls{mmwave} devices are mounted on the mobile robot. They are connected as AP (Initiator) and STA (Responder) by bouncing from the environment (self-sensing) and providing \gls{mmwave} sensing data in the form of \gls{ftm} and \gls{csi}. The LiDAR emits laser pulses and measures the time it takes for the pulses to bounce back after hitting objects in the environment. Odometry comes from the mobile robot base, and it represents data from motion sensors to estimate changes in position over time.

\subsection{Individual Point Estimations} For indoor mapping, we need a collection of individual point estimations from the surrounding environment. This process is called Point Cloud estimation, where a multitude of individual points are captured using \gls{mmwave} self-sensing and LiDAR.

\simpletitle{mmWave Individual point} To obtain an \gls{mmwave} individual 2D point estimate, \toolname is interested in getting the distance and the azimuth \gls{aoa} for the path that connects the two \gls{mmwave} devices via first-order reflection. Every time a path bounces off an obstacle, it gets attenuated by $5-10$~dB, depending on the material. 
As the mD-track algorithm provides the azimuth and elevation angles and the attenuation of every path, we can identify the strongest path. We then denote this path as $p_{f}$ and we select it as follows: 
\begin{equation}
    p_{f} = \argmax_p |\gamma_p|.
\end{equation}

The elevation and the azimuth angles for the $p_{f}$ path are $\alpha_{p_{f}}$ and $\beta_{p_{f}}$, respectively. We denote the estimated azimuth as $\hat{\beta} = \beta_{p_f}.$

Given the multi-modal data collection, the odometry, \gls{ftm} and \gls{csi} measurements are combined to interpolate a \gls{mmwave} individual point estimation. First, we compute the Responder's position from the robot's odometry and convert it to Cartesian coordinates. Next, as described in Sec.~\ref{sec:fundamentals}, we use the mD-track algorithm to extract the azimuth and elevation from the \gls{csi} estimate. 
To validate a measurement, we make sure the elevation in the \gls{csi} is close to zero. 
Finally, we compute the estimated distance between the two \gls{mmwave} devices via first-order reflection as:
\begin{equation}
\hat{d}=c \times \textrm{\gls{tof}}.    
\end{equation}

Inaccurate mmWave individual point estimations can occur in environments with narrow corridors or high obstacle density where the signal has more multipath components. As a result, the mmWave radios can experience bad channel conditions, interferences, and undesired reflections. However, signal classification based on received power can generally isolate the single reflection component.

\simpletitle{LiDAR Individual Point}
Building indoor maps with mobile robots and LiDAR is a state-of-the-art procedure where individual points from the environment are collected by emitting laser pulses. The time it takes for the light to bounce back in the obstacles, sweeping 360\textdegree angles at high frequency, gives an individual point per angle. For more information regarding LiDAR-based SLAM please refer to~\cite{lidar,r3}. Inaccurate LiDAR Individual point estimations can occur in environments with dense fog, smoke, excessive light, or translucent materials (e.g.,glass), leading to potential errors in the map generated by the SLAM algorithm. 

\simpletitle{Individual Point Selection} Since \toolname has two sources of individual points, it needs to filter the data for the point cloud creation and indoor map reconstruction. When different values are reported for mmWave and LiDAR sources, \toolname checks previous values and estimates which is the correct source of data.  
For example, when facing translucent materials, LiDAR reports distance as infinite, and mmWave reports a determined value. Once the system detects that anomaly, mmWave points are injected into the system and provide accurate information about the translucent obstacle location.

\subsection{SLAM Algorithm}
The last building block of \toolname is the SLAM algorithm that generates a Point Cloud from the individual points estimations and builds the indoor map. SLAM algorithms are out of the scope of this study so this building block relies on existing SLAM algorithms that combine the Point Clouds with robot odometry data to generate high-resolution grid maps. This implies that we don't do any modifications to the existing SLAM algorithms but instead, we modify the input data to include \gls{mmwave} sensing.


\section{Prototype Design and Implementation}
\label{sec:prototype}

To evaluate the feasibility and performance of \toolname, we developed a prototype that is illustrated in Fig.~\ref{fig:blocksys}--(left). We explain every part of the prototype in detail below.

\simpletitle{Mobile Robot Sensing} Kobuki Turtlebot 2, a very well-known research platform that can carry multiple components works as the mobile robot base. We mount an RPLiDAR A2 on top of it, a well-performant solution, compatible with ROS, and easy to use. Two \gls{mmwave} devices with the same orientation are deployed on top of the robot and separated by absorbing material that eliminates undesired paths, prevents interference, and increases the antenna's directivity. 
Although the two mmwave antenna lobes overlap only on the edges, the absorbing material prevents radiation and avoids the risk of direct communication. 

In addition, a Raspberry Pi is mounted on the robot to interconnect the system and send the raw data via WiFi to a remote server for offline analysis. While the mobile robot base and RPLiDAR are connected to the Raspberry Pi via USB, the \gls{mmwave} devices are connected via Ethernet. The control of the mobile robot and RPLiDAR are implemented on the Raspberry Pi using ROS~\cite{ros} which is a widespread robotic framework used in both industry and academia. Besides navigation, the Raspberry Pi is also responsible for odometry, LiDAR, and \gls{mmwave} data storage. As there are very few \gls{ftm} and \gls{csi} capable COTS \gls{mmwave} devices, we chose the MikroTik wAP 60G, shown in Fig.~\ref{fig:blocksys}--left. Since the COTS devices have an 802.11ad proprietary operating system with limited access, we use the implementation in~\cite{multiloc}, which ported OpenWRT to this platform that exposes \gls{ftm} and \gls{csi} measurements. Once the Lidar, \gls{ftm}, and \gls{csi} data are collected, the Raspberry Pi sends the collection to a server for offline individual point estimation.


\simpletitle{Individual Point Estimation} 
LiDAR individual points are collected using Software Development Kit (SDK) through a ROS node, as specified in RPLiDAR documentation.\footnote {Available online (accessed 11/07/2023): \url{http://wiki.ros.org/rplidar}}

For the \gls{mmwave} individual point estimation, we use a remote server where in MATLAB we implement the steps described in Sec.~\ref{sec:solution}. The remote server computes the Cartesian coordinates from the raw odometry data and the single reflection distance to an obstacle from the \gls{ftm} data. The raw \gls{csi} data is used to estimate the azimuth and elevation. The Cartesian robot coordinates, distance, elevation, and azimuth are used for the creation of individual point estimation. The MATLAB script also implements a filter based on previous data to detect if either LiDAR or mmWave data is not correct and decide the proper source of the point estimation.

\simpletitle{SLAM Algorithm} 
Hectormap~\cite{slam1} is used as an exemplary ranging-based \gls{slam} algorithm for both point cloud and grid map reconstruction. We used the ROS implementation of Hectormap to create a 2D occupancy grid map based on the filtered individual points estimated by LiDAR and \gls{mmwave}.

\section{Experimental Validation}




\label{sec:experimental}
This section evaluates the performance of \toolname prototype through practical experiments executed in a wide range of scenarios described in Sec.~\ref{subsec:testbed}. Specifically, Sec.~\ref{subsec:capabilities} elaborates on the capabilities and limitations of \toolname, and Sec.~\ref{subsec:gridmap} validates the grid map reconstruction achievable. 

\subsection{Testbed and Data Collection}
\label{subsec:testbed}

\begin{figure}
\centering%
\includegraphics[width=0.4\textwidth]{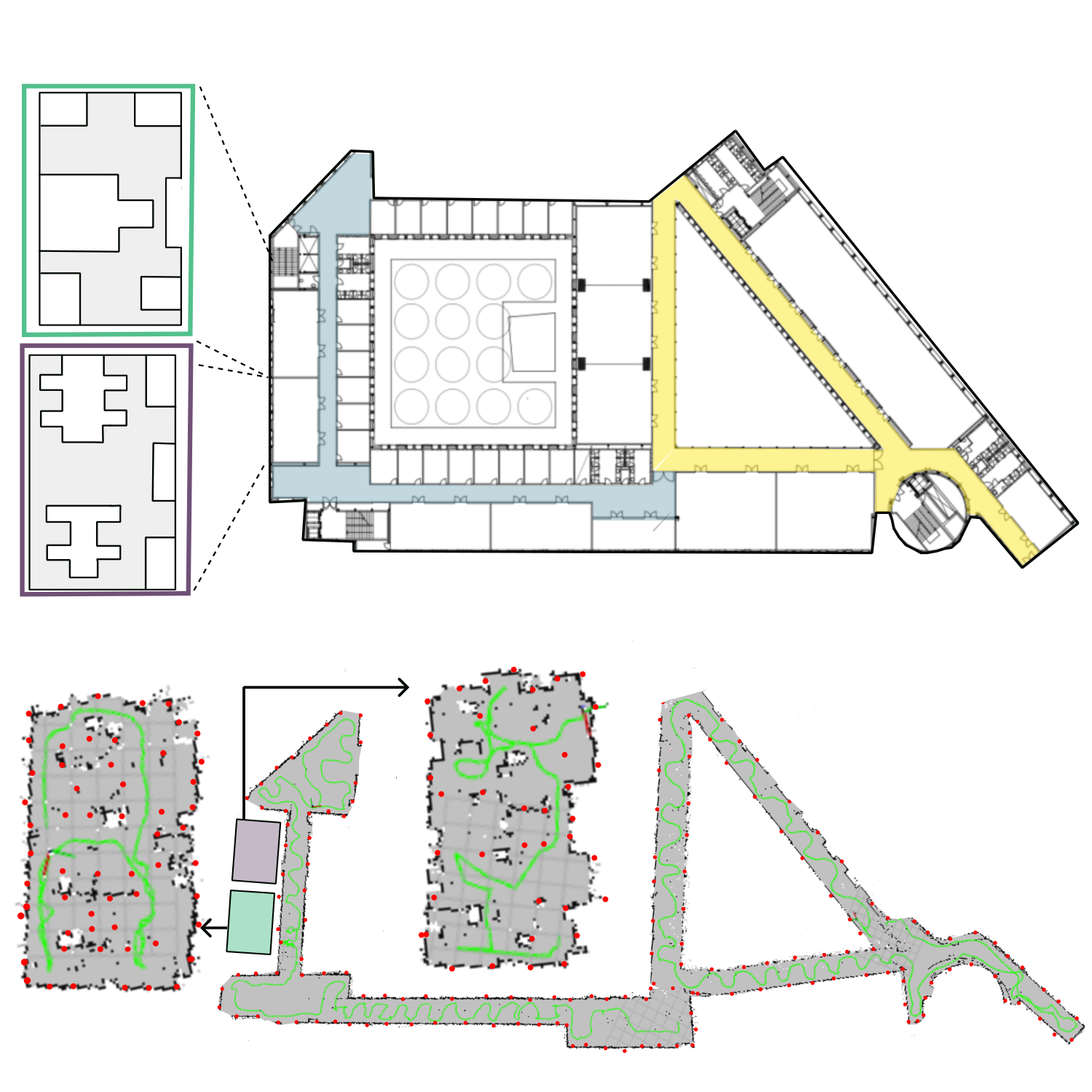}%
\vspace*{-2ex}%
\caption{Selected Testbed--(top) and \toolname map reconstruction--(bot)}%
\label{fig:mapuc3m}%
\vspace*{-3ex}%
\end{figure}

\simpletitle{Testbed}
In the experimental scenario for testing the \toolname, we map the floor of a UC3M building (see~Fig.~\ref{fig:mapuc3m})--(top), which presents a variety of different materials such as glass or brick walls. This scenario challenges \toolname to effectively handle reflective surfaces and obstacles, thus evaluating its robustness and reliability in real-world settings.

Furthermore, two individual labs are included in the testbed scenario. Mapping these specific labs provides an opportunity to test the \toolname tool's performance in more confined and specialized spaces. Fig.~\ref{fig:mapuc3m}--(bottom) presents the reconstruction of the UC3M floor with \toolname, which will be further elaborated in Subsec.~\ref{subsec:gridmap}.

\simpletitle{Data Collection} To collect the data needed for map reconstruction, we drove the mobile robot using a remote controller without setting any pre-defined route, following a random walk. We collected data from the \gls{mmwave} sensing (\gls{ftm} and \gls{csi}), LiDAR, and wheel odometry. During the data collection, the \gls{mmwave} devices on the robot are positioned facing toward the target object/material surface.

\subsection{waveSLAM Capabilities}
\label{subsec:capabilities}
To understand the capabilities and limitations of \toolname, we first conducted an analysis to elaborate on the accuracy of the estimations as a function of the distance between the robot and the walls, and the angle estimations.

\begin{figure}
\centering%
\subfloat[ToF extracted from FTM~\label{fig:capabilitiesdistance}]{%
    \includegraphics[width=.4475\columnwidth,keepaspectratio]{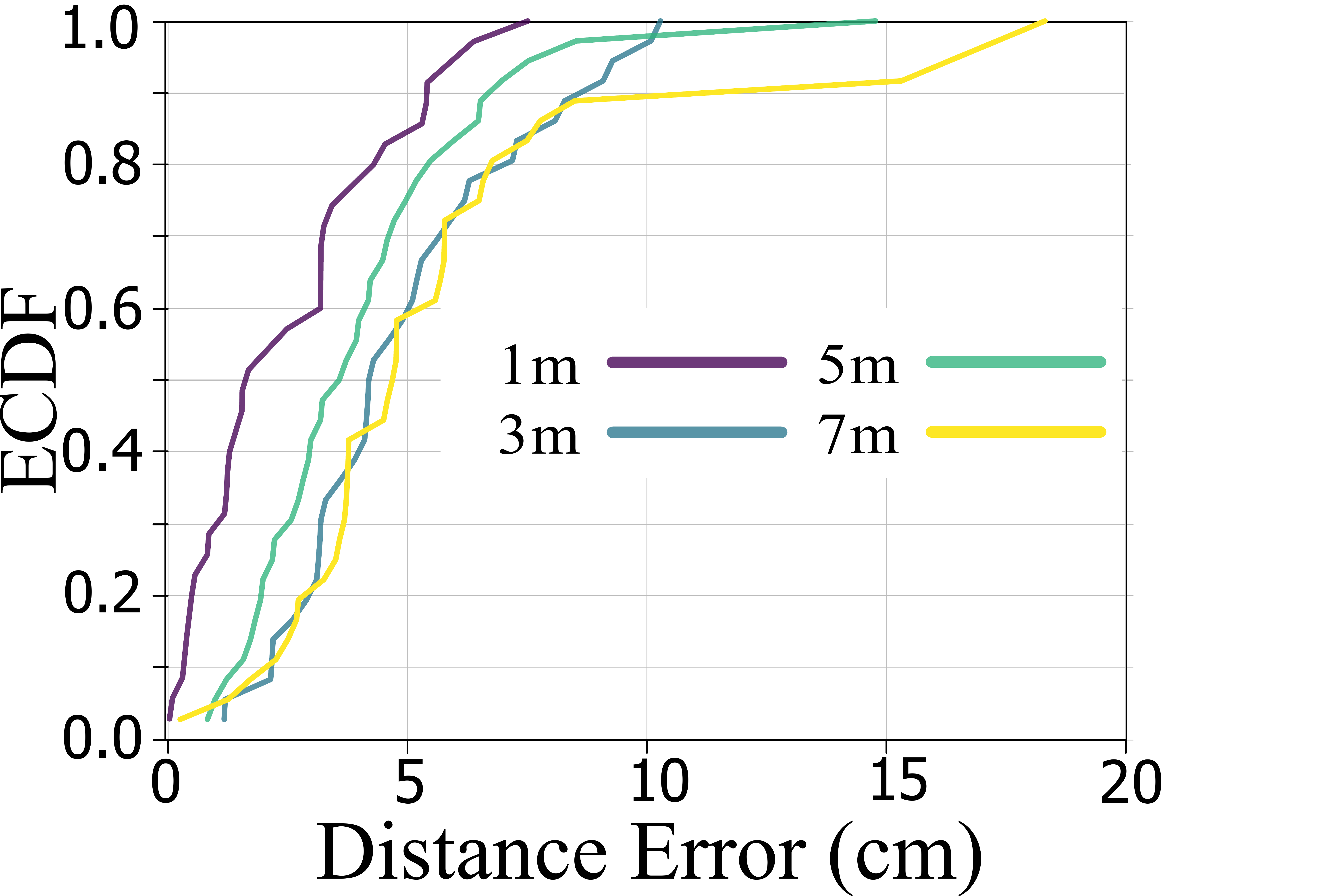}}%
~%
\subfloat[AoA estimations with URA~\label{fig:capabilitiesangle}]{%
    \includegraphics[width=.5425\columnwidth,keepaspectratio]{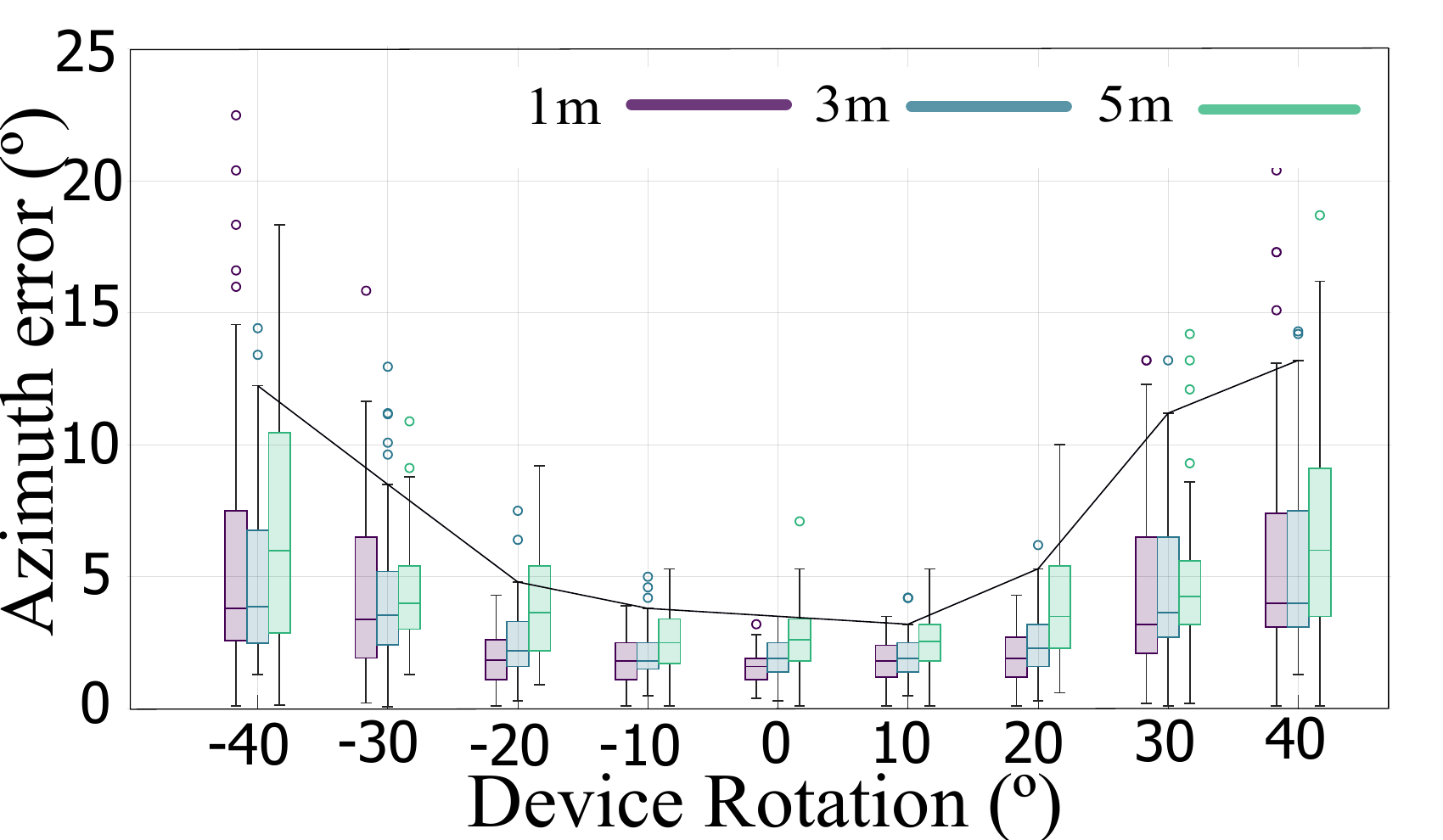}}%
\vspace*{-1ex}%
\caption{Capabilities of \gls{mmwave} devices}%
\label{fig:capabilities}
\vspace*{-3ex}%
\end{figure}

\simpletitle{Evaluation Method} We evaluate the estimated distances by placing our prototype facing a wall and increasing the distance between the robot and the wall, from $1$~m to $7$m, with a $2$~m step. Similarly, for angle estimations, we perform \gls{aoa} measurements for different angles ($-40$ to $40^{\circ}$) at different distances ($1$, $3$, and $5$~m away from the wall). For each distance and angle estimation, we perform $50$ measurements and generate the graphs shown in Fig.~\ref{fig:capabilities}. 

\simpletitle{Evaluation Results} Fig.~\ref{fig:capabilitiesdistance} shows the Empirical Cumulative Distribution Function (ECDF) of \toolname FTM distance error. We see that \toolname can achieve cm-level accuracy for all distances, keeping the distance error below $10$~cm for 80\% of the cases. Moreover, Fig.~\ref{fig:capabilitiesdistance} also shows that the FTM distance error increases as we increase the distance. The reason for this is that, when increasing the distance between the robot and the wall, the attenuation is higher and more sensitive to noise. In addition, the further you are from an obstacle, the more chances of getting a 2nd order or higher reflection the system has, resulting in bigger errors, as Fig.~\ref{fig:capabilitiesdistance} yellow line ($7$~m) shows. However, we can notice that even at 7m of distance the error remains at the centimeter level.

Fig.~\ref{fig:capabilitiesangle} shows a box plot of \toolname azimuth error. We can see that the angle estimation error of \toolname also increases as we increase the rotation angle of the robot, due to higher complexity in mD-track angle estimations. 
Distances above $7$~m and angles above $\pm 40^{\circ}$ the current prototype of \toolname has difficulty collecting data due to the increased number of reflections collected. These results demonstrate that the angle and distance estimations obtained from the \gls{cots} \gls{mmwave} devices can be accurate and useful for performing \gls{slam}. While angle estimations indicate that the navigation of the robot when performing \gls{slam} should always confront the object/material surface at angle orientations between $\pm 40^{\circ}$, the distance estimations demonstrate that we can achieve a range of up to $7$~m, a distance comparable with commercial LiDARs\footnote {Available online (accessed 04/04/2023): \url{https://www.slamtec.ai/home/rplidar_a2/}}.

\begin{figure*}
\centering%
\includegraphics[width=0.9\textwidth]{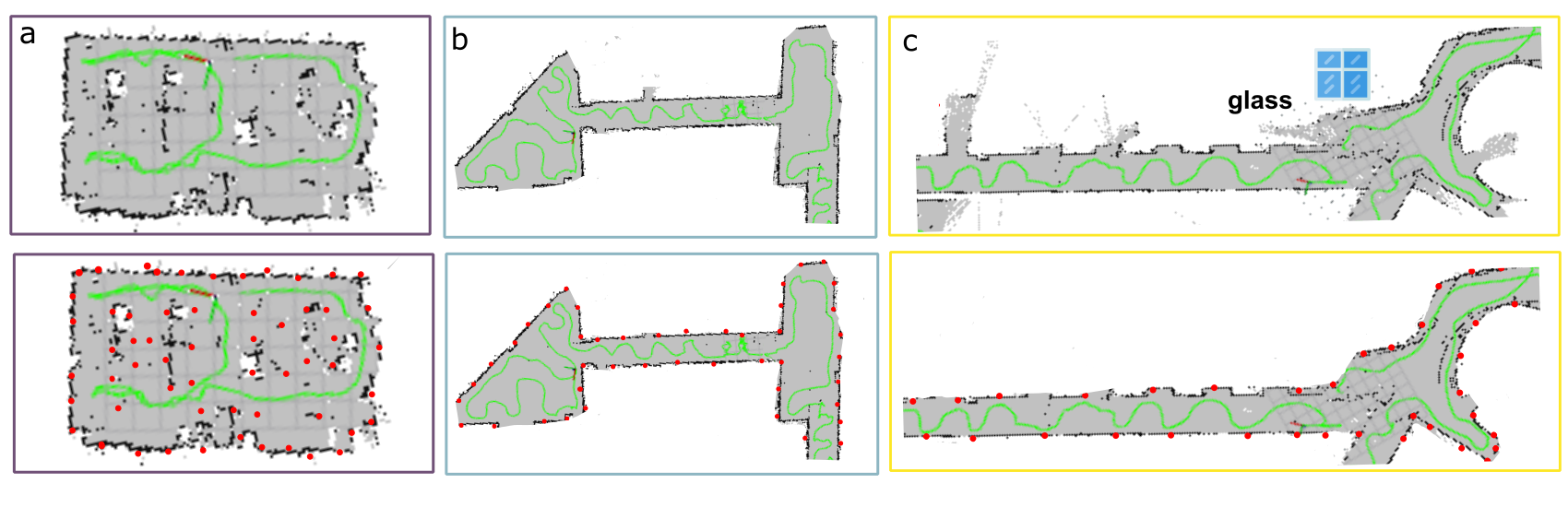}%
\vspace*{-2ex}%
\caption{LiDAR maps vs \toolname maps in a) closed lab b) dark corridor c) light corridor with partial glass walls}%
\label{fig:maps}%
\vspace*{-3ex}%
\end{figure*}

\begin{figure}
\includegraphics[width=.9\columnwidth,keepaspectratio]{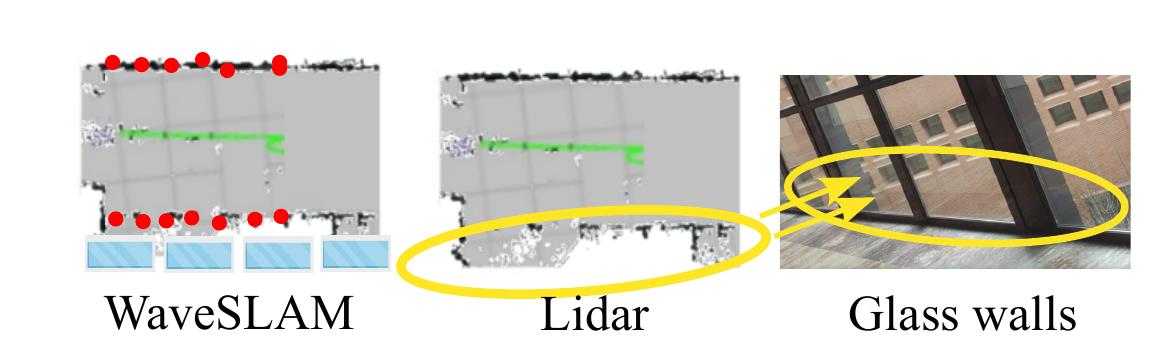}%
\vspace*{-1ex}%
\caption{Incorrect performance of LiDAR detecting glass walls}%
\label{fig:glasswall}%
\vspace*{-3ex}%
\end{figure}

\subsection{Grid Map Accuracy Performance}
\label{subsec:gridmap}
After understanding the capabilities of the \toolname prototype presented in Sec.~\ref{sec:prototype}, we proceed with the validation of the grid map reconstruction. 

\simpletitle{Evaluation Method} In this section we compare the grid map reconstruction accuracy of \toolname with existing state-of-the-art LiDAR-based indoor mapping. The robot was traversing the indoor corridors and at different collection points, it was rotated to a pose at which the \gls{mmwave} devices were parallel to the target object/material surface. With the LiDAR-generated map, we plot the \toolname points to empower the accuracy, especially in particular scenarios that are explained below.

\simpletitle{Evaluation Results} Fig.~\ref{fig:maps} zooms the reconstruction of three different parts of the map of UC3M previously shown in Fig.~\ref{fig:mapuc3m}--(bot), and compares the solution of just using RPLiDAR A2 against \toolname. In green, we indicate the route that the mobile robot follows, in black the LiDAR points while in red the mmWave points. From Fig.~\ref{fig:maps} (a) we notice that the accuracy in the lab room is similar, however, mmWave detects extra points as it is at a different height from LiDAR, gaining more spatial resolution to the map to locate obstacles. Fig.~\ref{fig:maps} (b) consists of a dark corridor, where LiDAR does not have problems mapping and mmWave points only compliment the LiDAR to validate the accuracy. Last, Fig.~\ref{fig:maps} (c) shows a light corridor with partial glass walls, turning into 1)light excess for a LiDAR when sunny and 2) the impossibility to detect translucent materials as the glass forming the walls. In this scenarios, \toolname manages to collect correct inputs since the glass wall is opaque to \gls{mmwave}, as shown in Fig.~\ref{fig:glasswall}.

\section{Related Work}
\label{sec:rel-works}
\simpletitle{Optical Sensing Solutions}
Mobile robots equipped with optical sensors (e.g., LiDAR, Camera) are the de-facto state-of-the-art for building high-resolution indoor maps. We refer the interested readers to the Optical Sensors \gls{slam} surveys~\cite{r1,r2}. Specific to building indoor maps with Optical Sensors,~\cite{r3} presented a robust \gls{slam} system based on monocular vision and LiDAR for robotic urban search and rescue. Conversely,~\cite{r4} presents an active \gls{slam} framework to create a collision-free trajectory, while ~\cite{r5} uses a forward-viewing monocular vision to create a map of the robot's surroundings. The use of RGB-D sensors in combination with computer vision algorithms has been considered in~\cite{r6,r7} to estimate the robot's position and map the surrounding environment. Mobile robots mounted with LiDARs and Cameras can build high-resolution indoor maps but they suffer serious performance degradations in situations with dust, lots of light, or poor illumination, limiting their use. \toolname overcomes these limitations by using the environment sensing information that comes from the robot low-cost \gls{cots} 60 GHz communication devices. 

\simpletitle{60GHz Sensing} 
In recent years, much research effort has been devoted to reconstructing the
indoor environment with devices operating at 60 GHz~\cite{mwave-1,mwave-2,mwave-3,mwave-4}. In particular, JADE~\cite{mwave-3} uses an innovative mapping algorithm to reconstruct the indoor environment but it assumes that it knows a-prior the information about the perfect signal path. The authors in~\cite{mwave-4} reconstruct the indoor environment by deploying multiple APs. In~\cite{mmwave-1} a ray-tracer is used to predict the quality of arbitrarily located links in a given environment structure but human intervention is needed to move the Tx/Rx pair to different locations in the indoor space. \cite{mmwave-2} makes use of an elaborated pose predictor, with limited application for commercial implementations. In addition,~\cite{mwave-4} and ~\cite{mmwave-1} use big and expensive SDRs with complex signal processing to achieve 60 GHz waveforms. 
The most similar work to \toolname is~\cite{zhou}, where mmRanger equips a pair of low-cost off-the-shelf mmWave radios which constantly sense the ambient environment using RSS-based solutions. However, their real-world deployment and practicality are limited to a micro-benchmark scenario. Recent works adapt also the radar technique~\cite{r10,r11,r12} to overcome the shortcomings of optical sensors. The mmWave radar-based technology has also been used in~\cite{r10}, where the authors propose, a low-cost, single-chip \gls{mmwave} radar-based indoor mapping system for assisting in emergency response scenarios. The \gls{slam} problem has also been addressed in~\cite{r11,r12}, where the authors introduce a new method for robotic mapping and navigation using \gls{mmwave} technology without prior knowledge of the environment. Still, the mmWave system in a radar-like configuration is just another sensor on the robot that is costly and not portable for industrial scenarios. In contrast, \toolname uses the LiDAR and the \gls{cots} \gls{mmwave} devices to reconstruct the occupancy grid map while the mobile robot travels in an indoor environment.

\vspace{1mm}

\toolname is positioned in the Joint Communication and Sensing category of solutions as it uses the COTS 60 GHz mmWave communication devices as an additional sensor for SLAM. We highlight that \toolname is not dependent on any expensive infrastructure or customized SDRs. In particular, \toolname exploits the existing state-of-the-art mobile robots that are equipped with LiDAR and mmWave radios to complement optical sensors SLAM solutions when they experience performance degradation. In recent years, using \gls{mmwave} radios to build indoor maps has gained significant attention and a variety of solutions have been proposed in the literacy~\cite{mwave-1,mwave-2,mwave-3,mwave-4}. However, the complexity behind such a \gls{mmwave} standalone solution limits its practical applicability in today's robotics systems. We strongly believe that solutions like \toolname can drastically decrease that complexity  by integrating the accurate \gls{mmwave} sensing data with existing optical sensors instead of completely substituting them.



\section{Discussion}
\label{sec:discussion}
\simpletitle{Higher-order Reflections}
In this work, we use the first-order reflection to generate a \gls{mmwave} individual point. However, \gls{csi} information considers second or higher-order reflections that, if properly reconstructed, can increase the point cloud density. This could be done by analyzing the \gls{aoa} and signal strength received to guess how many and where did the reflections bounced. We are working on advanced mapping methods to further improve the SLAM point cloud. 

\simpletitle{Standardization Activities} There are currently two parallel standardization initiatives, IEEE 802.11az~\cite{az} which covers the NextG positioning systems, and IEEE 802.11bf~\cite{bf} which deals with collaborative communication and sensing. While IEEE 802.11bf uses wireless signals received from IEEE 802.11 sensing capable STAs to determine the distance and angle estimations between them, IEEE 802.11az enables an STA to identify its relative position by extending the \gls{ftm} procedure to include not only distance but also angle estimations. A future version of \toolname can benefit from adopting 802.11az by obtaining the angle and distance estimations with only 802.11az FMT-ranging. In other words, this means that 802.11az has the potential to improve the data collection process of \toolname. 

\simpletitle{Improving Data Collection}
Limited by the customized version of the \gls{cots} MikroTik devices~\cite{multiloc} that were originally not designed to perform \gls{ftm} ranging and CSI estimations, \toolname can currently only collect a limited amount of measurements (e.g., 100). The MikroTik devices were suffering from overheating when \gls{ftm} and CSI measurements were invoked with higher frequency. Other robust commercial platforms that will be developed in the near future can be more suited as \gls{mmwave} devices for these types of mobile robots. Once the data collection frequency is improved, further trials need to be performed under diverse scenarios in order to understand the outliers as well as the sparse measurements returned by \gls{cots} \gls{mmwave} devices.

\simpletitle{\gls{mmwave}-based \gls{jcas} for SLAM}
This work is positioned in the \gls{jcas} category of solutions but it only uses the COTS 60 GHz mmWave devices for sensing. However, by incorporating scheduling mechanisms, our solution has the potential to efficiently allocate resources for both sensing and communications tasks, reusing sensing information to communicate with external \gls{sta}s achieving high-throughput directional beams. This introduces new challenges, such as finding proper scheduling algorithms not to degrade the sensing performance that we leave for in-depth study for future work.

\section{Conclusions}
\label{sec:conclusion}
Building indoor maps in low-visibility environments full of airborne obscurants is still a challenging problem today. To address this challenge, we presented \toolname, an efficient indoor mapping system that improves the performance of existing state-of-the-art optical SLAM solutions. In particular, \toolname uses a pair of \gls{cots} \gls{mmwave} radios to perform self-sensing and successfully map the indoor environment when the optical sensor fails. We implemented and validated \toolname with a mobile robot platform, showing its cm-level precision with errors below $22$~cm and $20^{\circ}$ in angle orientation across all configurations and settings.  


\bibliographystyle{IEEEtran}
\bibliography{IEEEabrv,IEEEsettings,ref}{}
\end{document}